\RequirePackage{lineno}
\documentclass[aps,floatfix,prc,twocolumn,showpacs,showkeys,amsmath,amssymb,superscriptaddress]{revtex4}

\usepackage{soul}
\usepackage{graphicx}	
\usepackage{xspace}     
\usepackage{amsmath}
\usepackage{hyperref}
\usepackage{color}
\usepackage{multirow}
\usepackage{booktabs,array}
\RequirePackage{lineno}
\newcommand{\Btwo}{$B_2$\xspace}
\newcommand{\Btwotrig}{$B_2^{trig}$\xspace}
\newcommand{\Btwoeve}{$B_2^{eve}$\xspace}

\newcommand{\dhcs}{di-hadron correlations\xspace}

\newcommand{\Dhcs}{Di-hadron correlations\xspace}

\newcommand{\pT}{$p_{\rm T}$\xspace}

\newcommand{\dphi}{$\Delta\phi$\xspace}
\newcommand{\dphirange}[1]{$|\Delta\phi|< $#1\xspace}
\newcommand{\detarange}[1]{$|\Delta\eta|< $#1\xspace}
\newcommand{\etarange}[1]{$|\eta|<$#1\xspace}
\newcommand{\deta}{$\Delta\eta$\xspace}

\newcommand{\sref}[1]{section~\ref{#1}}

\newcommand{\eref}[1]{equation~\ref{#1}}
\newcommand{\Sref}[1]{Section~\ref{#1}}
\newcommand{\Fref}[1]{Figure~\ref{#1}}

\newcommand{\Eref}[1]{Equation~\ref{#1}}

\newcommand{\pp}{$p$+$p$\xspace}

\newcommand{\sqrts}{$\sqrt{s}$}

\newcommand{\ns}{near-side\xspace}
\newcommand{\as}{away-side\xspace}

\newcommand{\ptassoc}{$p_T^{\mathrm{a}}$\xspace}
\newcommand{\pttrig}{$p_T^{\mathrm{t}}$\xspace}

\newcommand{\pttrigrange}[2]{#1 $< p_T^{\mathrm{t}} <$ #2~GeV/$c$\xspace}
\newcommand{\ptassocrange}[2]{#1 $< p_T^{\mathrm{a}} < $ #2~GeV/$c$\xspace}

\begin{document}

\title{Methods for separation of deuterons produced in the medium and
in jets in high energy collisions}

\author{Natasha Sharma} 
\affiliation{Department of Physics, Panjab University, Chandigarh, India-160014.}
\author{Tony Perez}
\affiliation{University of Tennessee, Knoxville, TN, USA-37996.}
\author{Andy Castro}
\affiliation{University of Tennessee, Knoxville, TN, USA-37996.}
\author{Lokesh Kumar} 
\affiliation{Department of Physics, Panjab University, Chandigarh, India-160014.}
\author{Christine Nattrass} 
\affiliation{University of Tennessee, Knoxville, TN, USA-37996.}

\date{\today}

\begin{abstract} 
Coalescence has long been used to describe the production of light
(anti-)nuclei in heavy ion collisions.  
The same underlying mechanism may also exist in jets when a proton and
a neutron are close enough in phase space to form a deuteron. 
 We model deuteron production in jets by applying an afterburner
 to protons and neutrons produced in PYTHIA for \pp collisions 
at a center of mass energy \sqrts\ = 7 TeV.  
PYTHIA provides a reasonable description of the proton spectra and the shape of the deuteron spectrum predicted by the afterburner is in agreement with the data. 
We show that the rise in the coalescence parameter \Btwo with momentum observed in data is consistent with coalescence in jets.  We show that \dhcs can be used to separate the contributions from the jet and the underlying event.  This model predicts that the conditional coalescence parameter in the jet-like correlation should be independent of the trigger momentum.
\end{abstract}

\maketitle

\section{Introduction}
The production of light (anti-) nuclei is of interest in high energy
particle collisions because of the insight that these measurements can
provide into particle production
mechanisms~\cite{Ahle:1999in,Armstrong:2000gz,Barrette:1999kq,Albergo:2002gi,Ambrosini:1997bf,Bearden:2002ta,Afanasev:2000ku,Anticic:2004yj,Anticic:2011ny,Anticic:2016ckv,Afanasiev:2007tv,Abelev:2008ab,Abelev:2009ae,Agakishiev:2011ib,Adam:2015vda}. Since
the binding energies of  
light (anti-) nuclei are on the order of a few MeV, they may be formed 
via coalescence of (anti-) 
nucleons in the later stages of evolution of the system~\cite{Gutbrod:1988gt,Butler:1963pp}. On the other hand, the description of light
(anti-) nuclei yields by thermal models might suggest their thermal
production~\cite{Adam:2015vda}. The recent results
from the STAR experiment on the coefficient of the second term of the Fourier decomposition of the azimuthal anisotropy, $v_2$, as a function of
transverse momentum of various nuclei show scaling with the number of
constituent nucleons~\cite{Adamczyk:2016gfs}. This behavior is
expected if light nuclei are formed by the coalescence of nucleons.

In the coalescence approach, the probability
of deuteron formation is related to the local density of constituent
nucleons as well as their velocities~\cite{Gutbrod:1988gt,Kapusta:1980zz}. The invariant yields
of light nuclei can be related to the yields of constituent nucleons
by 
\begin{eqnarray}
	{E_A}\frac{{{d^3}{N_A}}}{{{d^3}{p_A}}} = {B_A}{\left(
            {{E_p}\frac{{{d^3}{N_p}}}{{{d^3}{p_p}}}} \right)^Z}{\left(
            {{E_n}\frac{{{d^3}{N_n}}}{{{d^3}{p_n}}}} \right)^{A - Z}}
        \nonumber \\ \approx {B_A}{\left( {{E_p}\frac{{{d^3}{N_p}}}{{{d^3}{p_p}}}} \right)^A},
\label{eq1}
\end{eqnarray}
where $N_A$, $N_p$, and $N_n$ represent the yields of a given nucleus,
constituent protons, and consituent neutrons, respectively, and $p_A$, 
$p_p$, and $p_n$ are their momenta such that $p_p = p_n = \frac{p_A}{A}$. $A$ and $Z$ are the atomic mass
number and atomic number, respectively.  The coalescence parameter $B_A$ reflects the probability of
nucleon coalescence. Since the coalescence is expected to happen at a later stage of
evolution of the system, the coalescence parameter $B_A$ can provide
information on the freeze-out correlation volume~\cite{Gutbrod:1988gt}, the effective
volume of the nuclear matter at freeze-out, i.e., $B_A \propto
V_{\mathrm{eff}}^{1-A}$. The coalescence
parameter $B_2$ in heavy-ion collisions decreases with increasing collision energy, consistent with
increasing source volume. Thus, $B_2$ measurements are also related to the
homogeneity volume from HBT~\cite{Adare:2014qvs,Scheibl:1998tk}.

Deuteron production has been measured recently in $pp$
collisions~\cite{Adam:2015vda,Acharya:2017fvb}, an interesting data set to study production through coalescence.
Measurements of anti-nuclei in $pp$ collisions are helpful in searches for dark matter~\cite{Blum:2017qnn}.
The ALICE experiment has measured the multiplicity dependence
of the $d/p$ ratio in $pp$, $p$-Pb, and Pb--Pb
collisions~\cite{Sharma:2016vpz}. The ratio increases with multiplicity from $pp$
to $p$--Pb collisions, approaching the level in Pb--Pb collisions
where it remains constant as a function of multiplicity. The full
range of data could not be fully explained by either a coalescence or a thermal model~\cite{Sharma:2016vpz}.

In this paper we study the coalescence mechanism for deuteron
production in $pp$ collisions in the Monte Carlo event generator
PYTHIA~\cite{Sjostrand:2006za}.
The PYTHIA model does not generate deuterons so we use the coalescence
mechanism as an afterburner for their production from the protons and
neutrons generated in PYTHIA. We use the same approach as was used in
Ref.~\cite{Chen:2003ava}.  

\Sref{Sec:Afterburner} describes the afterburner and
\sref{Sec:Spectra} compares the proton and deuteron spectra in this
model to data, demonstrating that this approach generally agrees with
the data.  We propose the use of \dhcs for separating the production
of deuterons in jets and in the bulk in \Sref{Sec:Correlations}.  This
would allow a data-driven approach to testing production mechanisms.
We summarize in \sref{Sec:Conclusions}. 
\section{PYTHIA with coalescence afterburner}\label{Sec:Afterburner}
PYTHIA is a Monte Carlo model which can simulate \pp
collisions~\cite{Sjostrand:2006za} and has been tuned to several
measurements so that it provides a reasonable description of the
data~\cite{Skands:2010ak}.  PYTHIA includes multiparton interactions
in order to describe the production of low and intermediate particle
production.  It includes the production of jets, minijets, and
some resonances but does not include correlations from mechanisms such
as hydrodynamical flow.  At high momenta, particle production in PYTHIA
is dominated by jets.  It therefore can describe jet production.
Since PYTHIA uses the Lund string model for parton fragmentation and has no mechanism for the production of deuterons, it does not predict the production of deuterons in \pp collisions. Hence, deuterons are not formed by default in PYTHIA and we require an afterburner to coalesce protons and neutrons into deuterons.

\begin{figure}
\begin{center}
\rotatebox{0}{\resizebox{\columnwidth}{!}{
        \includegraphics{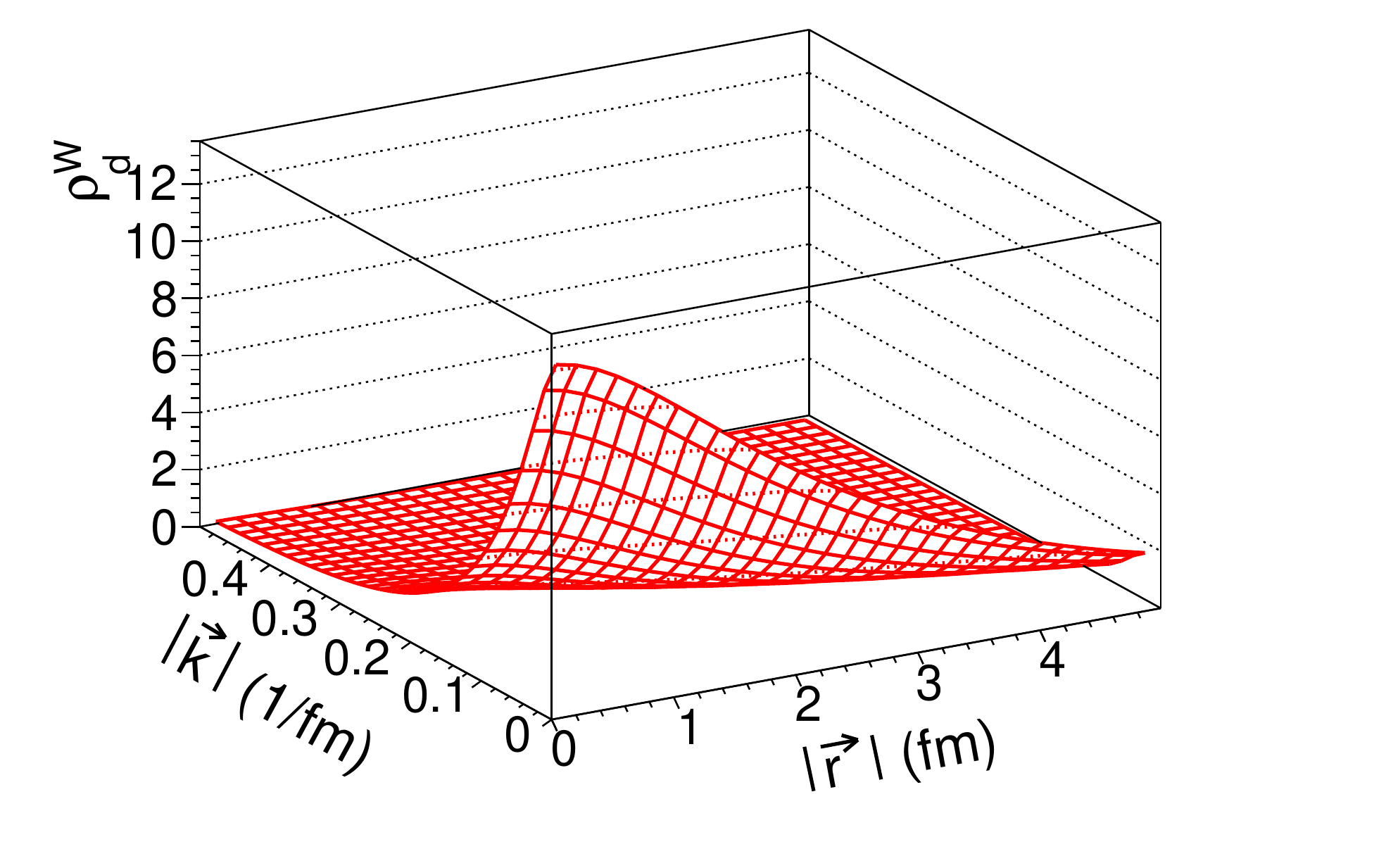}
}}\caption{The deuteron wave function $\rho^{W}_{d}$ defined in \eref{Eq:Wigner}
as a function of magnitude of relative position $\vec{r}$ and relative momentum $\vec{k}$ of protons and
neutrons for $\vec{r}\cdot \vec{k} = 0$.}\label{Fig:Wigner}
\end{center}
\end{figure}

The coalescence of protons and neutrons close in phase space has been used
to describe the production of deuterons in nuclear
collisions~\cite{Gutbrod:1988gt,Butler:1963pp,Kapusta:1980zz}.  
We use an afterburner for the production of deuterons
as described in Ref.~\cite{Chen:2003ava}.  
In this work,
the Wigner phase-space density for deuterons is obtained
from the Hulth\'en wave function which is
expressed in terms of the sum of 15 Gaussian wave functions
\begin{align}\label{Eq:Hulthen}
 &\phi(r) = \sum_{i=1}^{15} c_i (\frac{2w_i}{\pi})^{3/4} \exp(-w_ir^2),
\end{align}
where the coefficient $c_i$ and the width parameter $w_i$ are determined
by least square fit, and are 
given in Table~\ref{Tab:Wigner}. It is observed that the sum of 15
Gaussian wave functions reproduced the exact Hulth\'en wave function
both in coordinate and momentum spaces. The
deuteron wave function $\rho^{W}_{d}(\vec{r},\vec{k})$ can then be described analytically by the Wigner phase space density
\begin{align}\label{Eq:Wigner}
 &\rho^{W}_{d}(\vec{r},\vec{k}) = 8 \sum_{i=1}^{15} c_i^2 e^{-2 w_i r^2 - k^2/2 w_i}
 +\\ \nonumber
  16 \sum_{i>j}^{15} c_i c_j & (\frac{4 w_i w_j}{(w_i+w_j)^2})^{3/4} e^{\frac{-4w_i w_j r^2 - k^2}{w_i+w_j}} cos(2 \frac{w_i-w_j}{w_i+w_j} \vec{r} \cdot \vec{k}),
\end{align}
\noindent where $\vec{r}$ is the relative position of the proton and
neutron and $\vec{k}$ is their relative momentum. 
We evaluate \eref{Eq:Wigner} for the positions and momenta of final state protons and
neutrons in PYTHIA.  
Figure~\ref{Fig:Wigner} shows \eref{Eq:Wigner}
as a function of the relative positions and momenta of protons and
neutrons. It shows that the probability of deuteron formation through the
coalescence of protons and 
neutrons is higher if they are closer in momentum and coordinate space. The
probability decreases with increasing relative position and
momentum. Using protons and neutrons produced in PYTHIA and 
the afterburner described above, we obtain the deuterons from
coalescence of protons and neutrons. 

We use the PYTHIA~\cite{Sjostrand:2006za} Perugia 2011 tune~\cite{Skands:2010ak}  for the coalescence afterburner for  deuteron production. A total of 1.1 billion \pp collisions, including diffractive events, were simulated and unstable particles were forced to decay.  The number of deuterons is likely overestimated because PYTHIA assigns the primary vertex as the origin of primary particles.  This is a reasonable estimate for many purposes, but may overestimate spatial correlations between protons and neutrons.
We nevertheless expect that this method can describe the
qualitative behavior of deuteron production in experimental data and allow tests of experimental techniques for investigating deuteron production.

\begin{table}
\begin{center}
\caption{Coefficients describing the Wigner function in \eref{Eq:Wigner}}
\label{Tab:Wigner}
\begin{tabular}{c c c}
\hline \hline
$i$ & $c_i$ & $w_i$ (1/fm$^2$)\\ \hline
1 & 3.49665E-1 & 1.57957E-2 \\
2 & 1.85419E-1 & 3.94293E-2 \\
3 & 1.72279E-1 & 8.99793E-2 \\
4 & 4.62152E-2 & 9.75943E-2 \\
5 & 1.49458E-1 & 1.80117E-1 \\
6 & 7.74205E-2 & 1.93353E-1 \\
7 & 1.48268E-4 & 1.99811E-1 \\
8 & 7.35549E-3 & 2.17921E-1 \\
9 & 4.89047E-2 & 2.89902E-1 \\
10 & 4.19816E-2 & 4.70739E-1 \\
11 & 1.72670E-2 & 4.89604E-1 \\
12 & 1.06294E-1 & 9.27621E-1 \\
13 & 2.51462E-4 & 1.98822E+0 \\
14 & 3.22947E-2 & 2.59243E+0 \\
15 & 1.15826E-2 & 1.44639E+1 \\ \hline \hline
\end{tabular}
\end{center}
\end{table}
 
\section{Results}
\subsection{Spectra}
\begin{figure}
\begin{center}
\rotatebox{0}{\resizebox{\columnwidth}{!}{
        \includegraphics{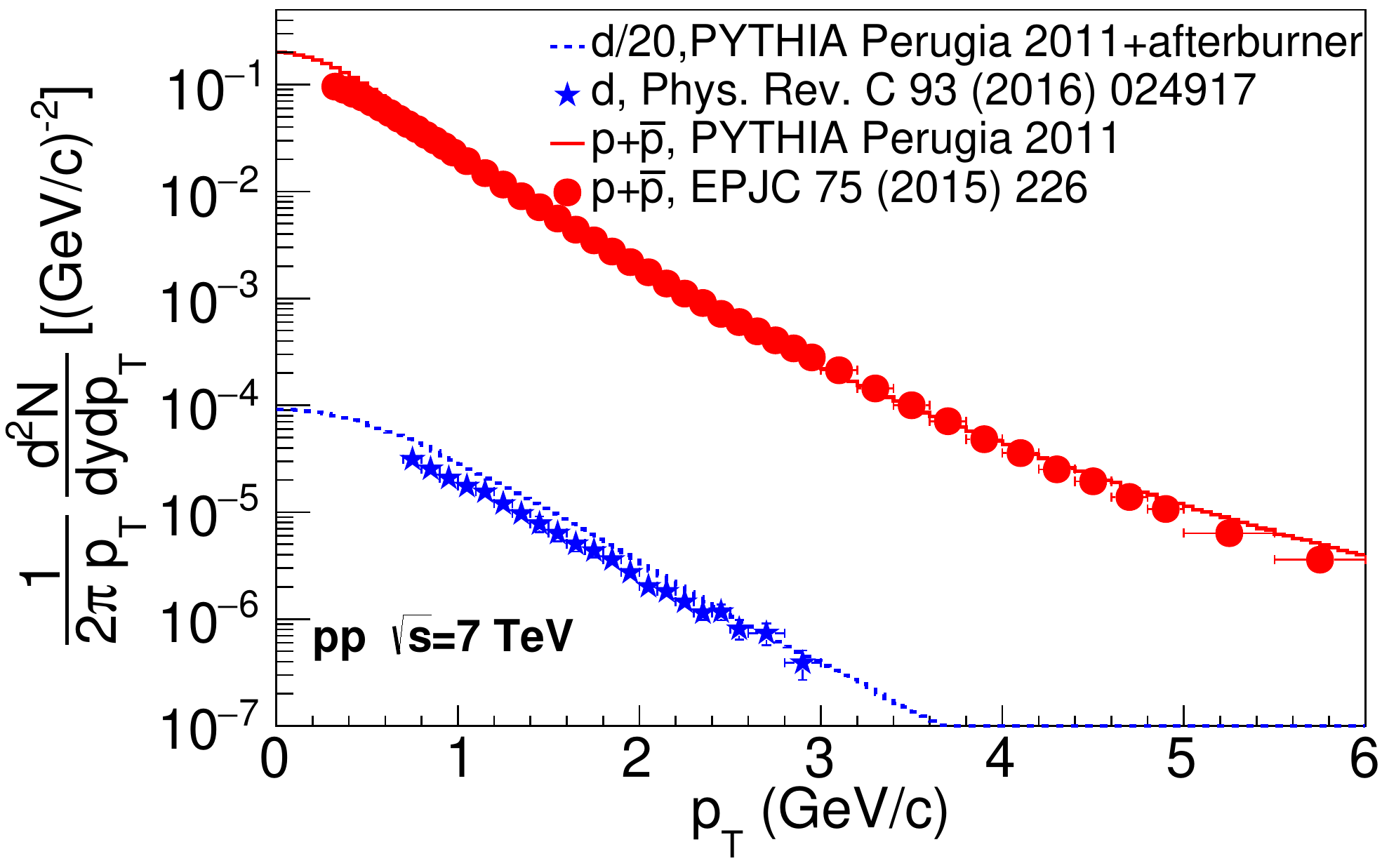}
}}\caption{Proton spectra from~\cite{Adam:2015qaa} and PYTHIA~\cite{Sjostrand:2006za} Perugia 2011 tune~\cite{Skands:2010ak} and deuteron spectra from~\cite{Adam:2015vda} and PYTHIA Perugia 2011 with the coalescence afterburner described in the text in \pp collisions at \sqrts = 7 TeV.}\label{Fig:Spectrapp}
\end{center}
\end{figure}

\Fref{Fig:Spectrapp} shows the deuteron spectrum from PYTHIA with the coalescence afterburner compared to measurements~\cite{Adam:2015vda,Acharya:2017fvb}. The deuteron spectrum from PYTHIA is scaled by a factor of 20.  As explained in \Sref{Sec:Afterburner}, PYTHIA likely overestimates spatial correlations between protons and neutrons, leading to an overestimate of the number of deuterons produced. The deuteron spectrum from the model has a similar shape to that observed in data. To see how the proton spectrum from PYTHIA compares to the data, we also plot the $p$+$\bar{p}$ spectrum from PYTHIA with the same from data~\cite{Adam:2015qaa}. PYTHIA gives the correct shape for both protons and deuterons.

\begin{figure}
\begin{center}
\rotatebox{0}{\resizebox{\columnwidth}{!}{
        \includegraphics{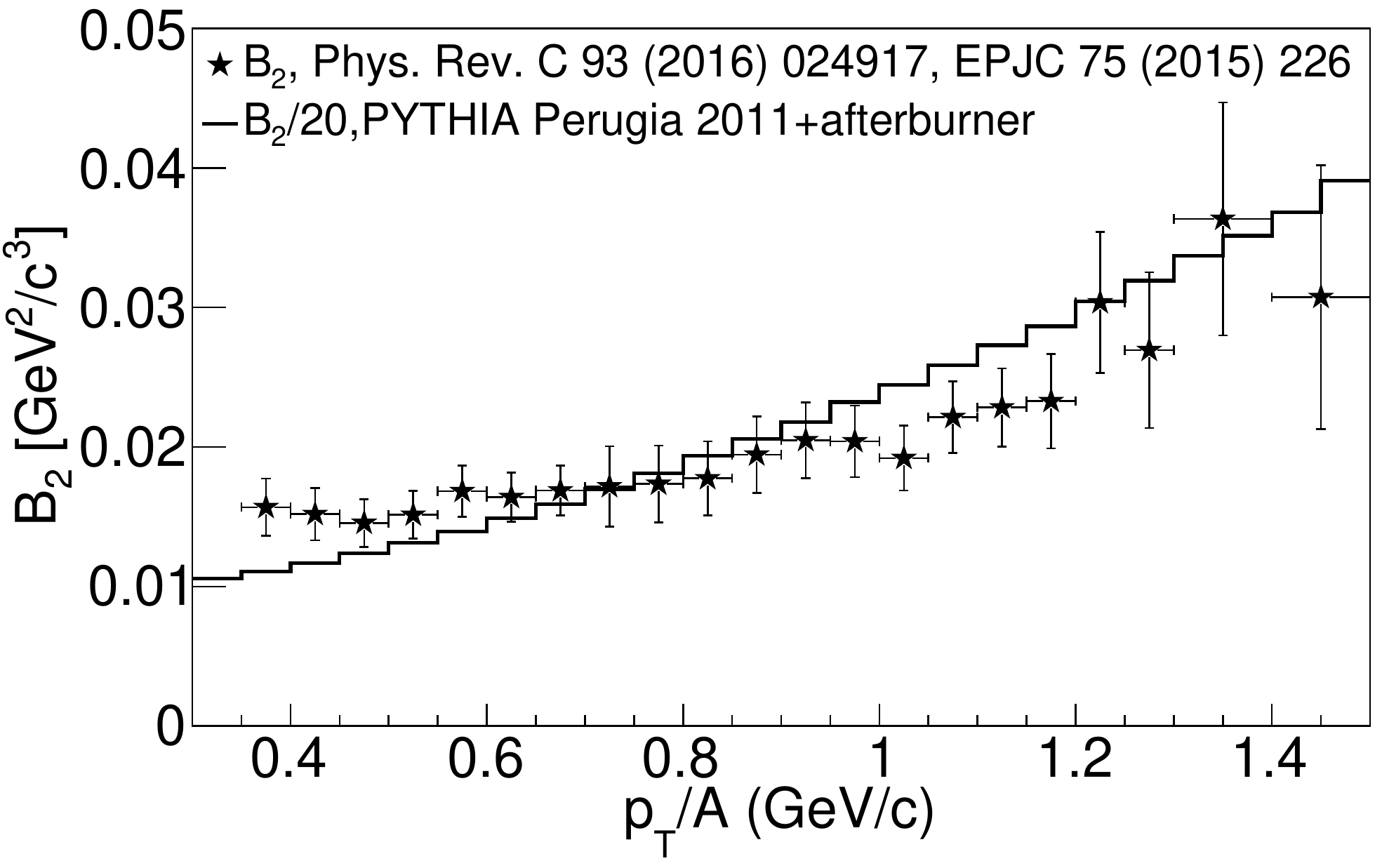}
}}\caption{\Btwo calculated from proton~\cite{Adam:2015qaa} and deuteron~\cite{Adam:2015vda} spectra and from PYTHIA~\cite{Sjostrand:2006za} Perugia 2011 tune~\cite{Skands:2010ak} with the coalescence afterburner described in the text in \pp collisions at \sqrts = 7 TeV.  
}\label{Fig:B2PYTHIA}
\end{center}
\end{figure}

\Fref{Fig:B2PYTHIA} shows \Btwo calculated using \eref{eq1}  from proton~\cite{Adam:2015qaa} and deuteron~\cite{Adam:2015vda} spectra and from the PYTHIA~\cite{Sjostrand:2006za} Perugia 2011 tune~\cite{Skands:2010ak} with the coalescence afterburner in \pp collisions at \sqrts = 7 TeV.  
The \Btwo from PYTHIA+afterburner overestimates deuterons and has been scaled by 20. 
The \Btwo from data and PYTHIA+afterburner show similar behavior, increasing as a function of \pT with similar slopes.  

In heavy ion collisions, \Btwo also shows an increase with increasing \pT~\cite{Adam:2015vda}.  It is expected that \Btwo as a function of \pT would be flat if the deuterons are formed via simple coalescence~\cite{Adam:2015vda}. The rise in \Btwo as a function of \pT in data is suggested to be due to flow~\cite{Adam:2015vda} and/or hard scattering~\cite{Liu:2006my,Acharya:2017fvb}. 
The PYTHIA model does not include flow so the rise in \Btwo as a function of \pT in the PYTHIA+afterburner model can be attributed to hard scattering.   Moreover, particle production at high momenta in PYTHIA is dominated by jets and mini-jets.
The contribution to deuteron production through coalescence in jets can by studied  in jets and the contribution from soft processes can be studied in the underlying event.
It is therefore interesting to study the behavior of \Btwo versus \pT for jets and the underlying event separately. 

 \label{Sec:Spectra}

\subsection{\Dhcs}\label{Sec:Correlations}
\begin{figure*}
\begin{center}
\rotatebox{0}{\resizebox{\textwidth}{!}{
        \includegraphics{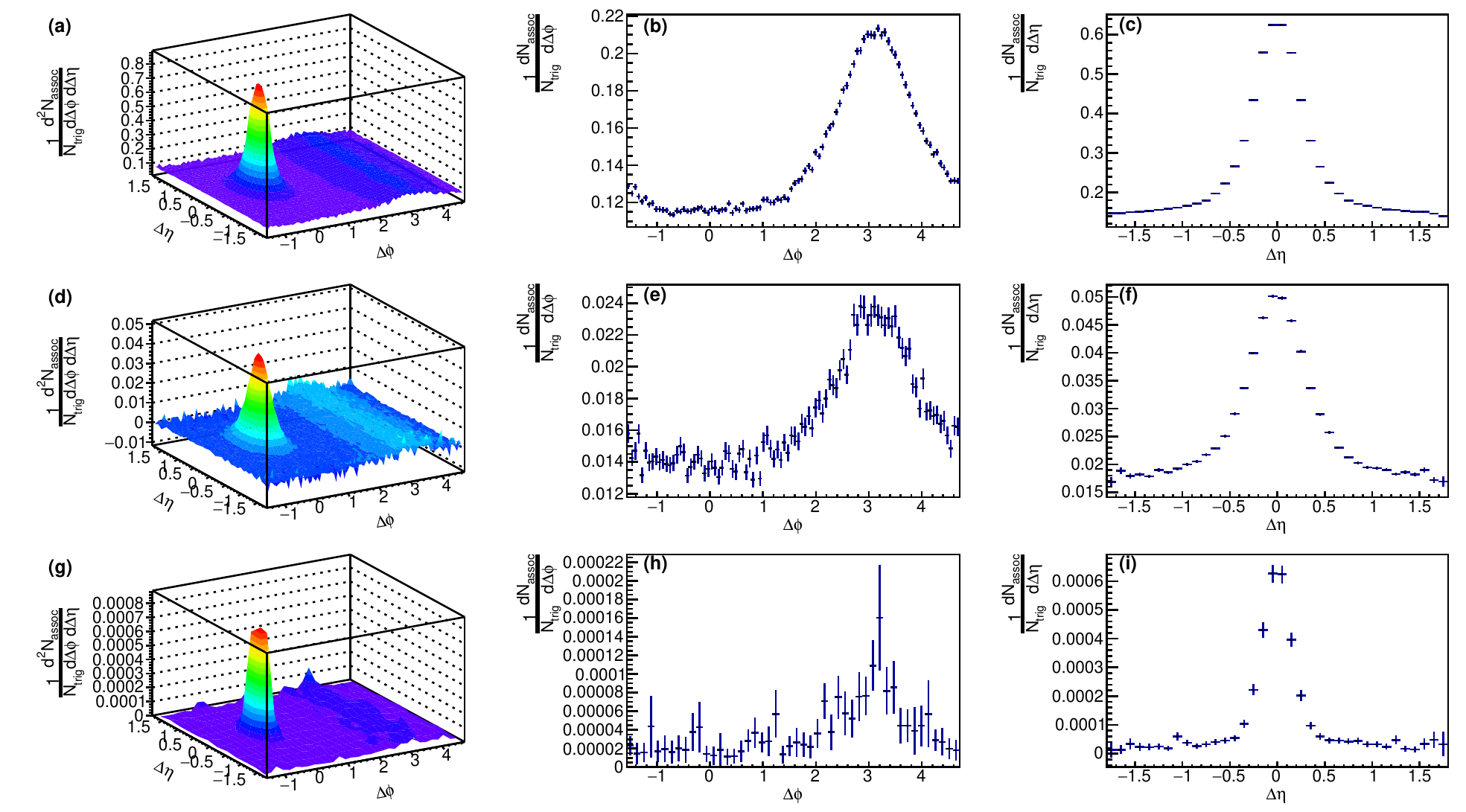}
}}
  \caption{\Dhcs for trigger momenta \pttrigrange{5}{7} within pseudorapidities \etarange{0.9} and associated particles within \etarange{0.9} with momenta \ptassocrange{2.0}{3.0} in \pp collisions at \sqrts\ = 7 TeV in PYTHIA~\cite{Sjostrand:2006za}.  The signal is normalized by the number of trigger particles and corrected for the acceptance as described in the text.  Correlation function in \dphi and \deta is shown for associated charged hadrons (a), associated protons (d), and associated deuterons (g).  Correlation function in \dphi for $1.0< |\Delta\eta|< 1.8$ is shown for associated charged hadrons (b), associated protons (e), and associated deuterons (h).  Correlation function in \deta for $-\frac{\pi}{2}< |\Delta\phi|< \frac{\pi}{2}$ is shown for associated charged hadrons (c), associated protons (f), and associated deuterons (i).}\label{Fig:SamplePlots}
\end{center}
\end{figure*}

We propose disentangling the contributions from the underlying event and in jets using \dhcs, which are frequently used to measure the production of jets in heavy ion collisions without the need for full jet reconstruction and allow separation between contributions from jets and from flow.  

A trigger particle is selected by its high momentum, \pttrig, and the distribution of particles is measured in azimuth $\Delta\phi = \phi^t - \phi^a$ and pseudorapidity $\Delta\eta = \eta^t - \eta^a$ relative to that trigger particle.  
Here the superscripts $t$ and $a$ are used for the trigger and associated particles, respectively.
We restrict reconstructed particles to \etarange{0.9}, a range accessible by the ALICE experiment, resulting in a trivial acceptance effect.  We correct for this by dividing by $a(\Delta\eta) = 1 - \frac{1}{1.8} \Delta \eta$.  A sample correlation is shown in 
Fig.~\ref{Fig:SamplePlots}
for charged hadron ($\pi^{\pm}$, $K^{\pm}$, $p$, $\bar{p}$) trigger particles with charged hadron associated particles, for charged hadron trigger particles with associated protons, and for charged hadron trigger particles with associated deuterons.  A clear peak is seen near $\Delta \phi \approx 0$ and $\Delta \eta \approx 0$, referred to as the \ns.  This peak is narrow in both \dphi and \deta and contains particles from the same jet as the trigger particle.  There is an additional peak from the partner jet at approximately $\Delta \phi = \pi$, called the \as.  This peak is narrow in azimuth but broad in pseudorapidity due to the difference between the center of momentum frame of the hard scattered partons and the incoming protons.

Even in PYTHIA, there is some contribution from the underlying event.  \Dhcs contain contributions where both the trigger and associated particles are from the same jet ($J-J$), where the trigger particle is from a jet but the associated particle is not from the same jet ($J-B$), and where neither the trigger or associated particles are from a jet ($B-B$).  At sufficiently high momenta, contributions from $B-B$ to the correlation function are negligible in \pp collisions.  In PYTHIA contributions from $J-B$ can include either associated particles from the underlying event or associated particles produced by a different hard scattering from the trigger particle.  Our goal here is to clearly identify deuteron production in jets, $J-J$, requiring a background subtraction.

There are several approaches to background subtraction in \dhcs~\cite{Sharma:2015qra}.  For simplicity, we focus on the \ns.  \Fref{Fig:SamplePlots} shows the correlation function in \dphi for $1.0< |\Delta\eta|< 1.8$ for each combination of trigger and associated particles, demonstrating that contributions of jet-like correlations on the \ns to the correlation function are negligible in this range.  The correlation function in \deta on the \ns, $-\frac{\pi}{2}< |\Delta\phi|< \frac{\pi}{2}$, is also shown.  We estimate the background by fitting this correlation function with a constant over the range $1.0< |\Delta\eta|< 1.8$.  This approach works in heavy ion collisions for subtracting the flow-modulated background as well~\cite{Abelev:2009af,Abelev:2009ah,Agakishiev:2011st}.  

\begin{figure}
\begin{center}
\rotatebox{0}{\resizebox{\columnwidth}{!}{
        \includegraphics{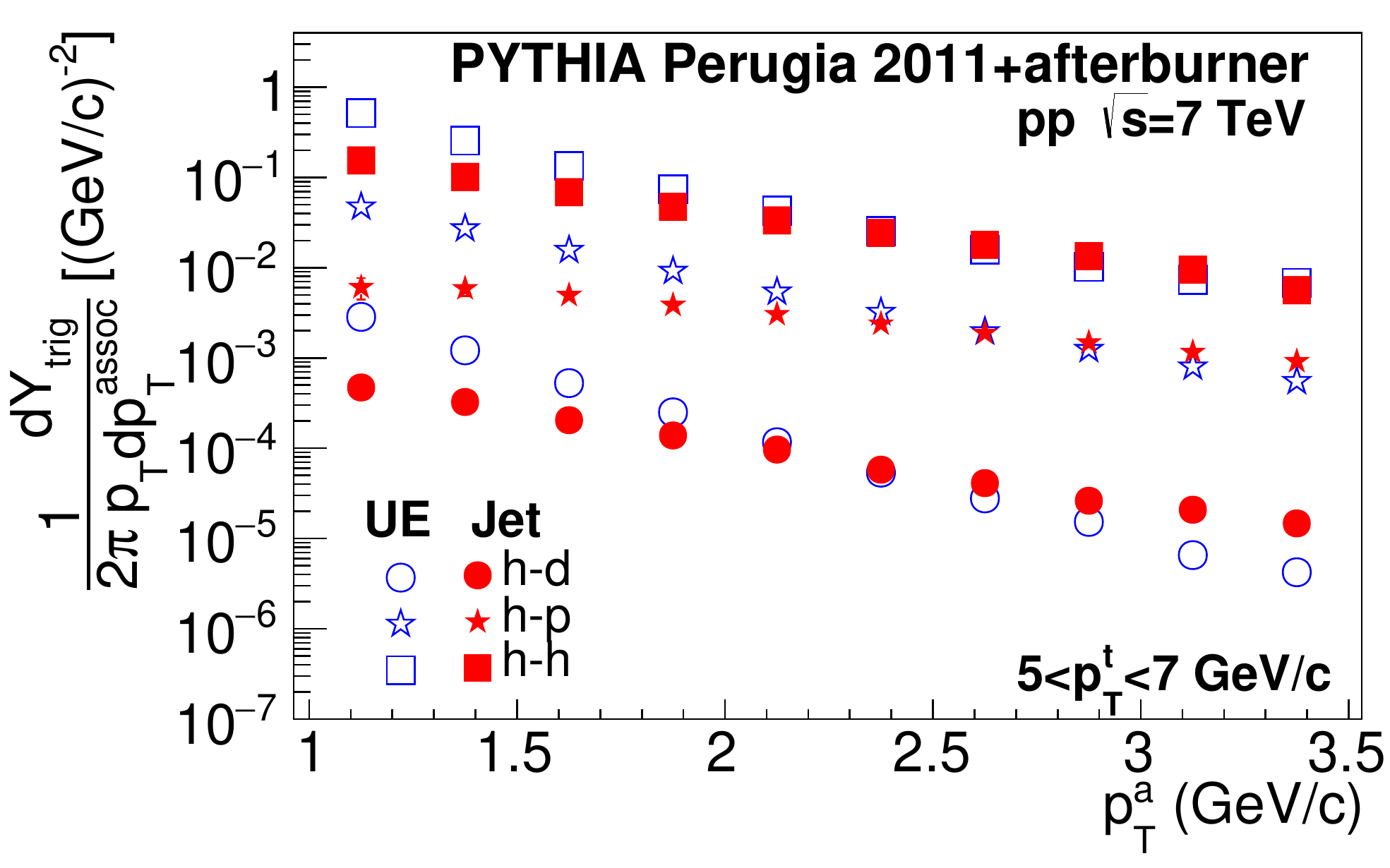}
}}\caption{Conditional spectra $\frac{1}{2\pi p_{T}}\frac{dY_{trig}}{dp_{T}}$ for \detarange{1.8} and \dphirange{$\frac{\pi}{2}$} as a function of \ptassoc for associated charged hadrons (h-h), associated protons (h-p), and associated deuterons (h-d) for \pttrigrange{5}{7} within pseudorapidities \etarange{0.9} and associated particles within \etarange{0.9} in \pp collisions at \sqrts\ = 7 TeV in PYTHIA~\cite{Sjostrand:2006za}.}\label{Fig:ConditionalSpectra}
\end{center}
\end{figure}

The conditional yield can be calculated as
\begin{equation}
 Y_{trig} = \frac{1}{N_{trig}}\int_{-1.8}^{1.8} \int_{-\frac{\pi}{2}}^{\frac{\pi}{2}} \frac{d^2N_{assoc}}{d\Delta\eta d\Delta\phi} d\Delta\phi d\Delta\eta \label{Eq:Yield}
\end{equation}
either for the $J-J$ signal or for the background and the conditional associated spectrum $\frac{1}{2\pi p_{T}}\frac{dY_{trig}}{dp_{T}}$ can be calculated.  \Eref{Eq:Yield} gives the yield normalized per trigger particle, the conventional normalization for \dhcs.  In the case when the number of trigger particles which are not from hard processes is negligible, the yield is then the number of associated particles per jet.  Yields normalized by the number of trigger particles should be comparable for different systems.  When the yield is normalized by the number of events, $Y_{eve}$, it is a measure of the number of particles produced by jets per event, which can then be compared to the inclusive particle spectra.  \Fref{Fig:ConditionalSpectra} shows the conditional associated spectra for the signal and the background for \pttrigrange{5}{7} in \pp collisions simulated with PYTHIA+afterburner.

\begin{figure}
\begin{center}
\rotatebox{0}{\resizebox{\columnwidth}{!}{
        \includegraphics{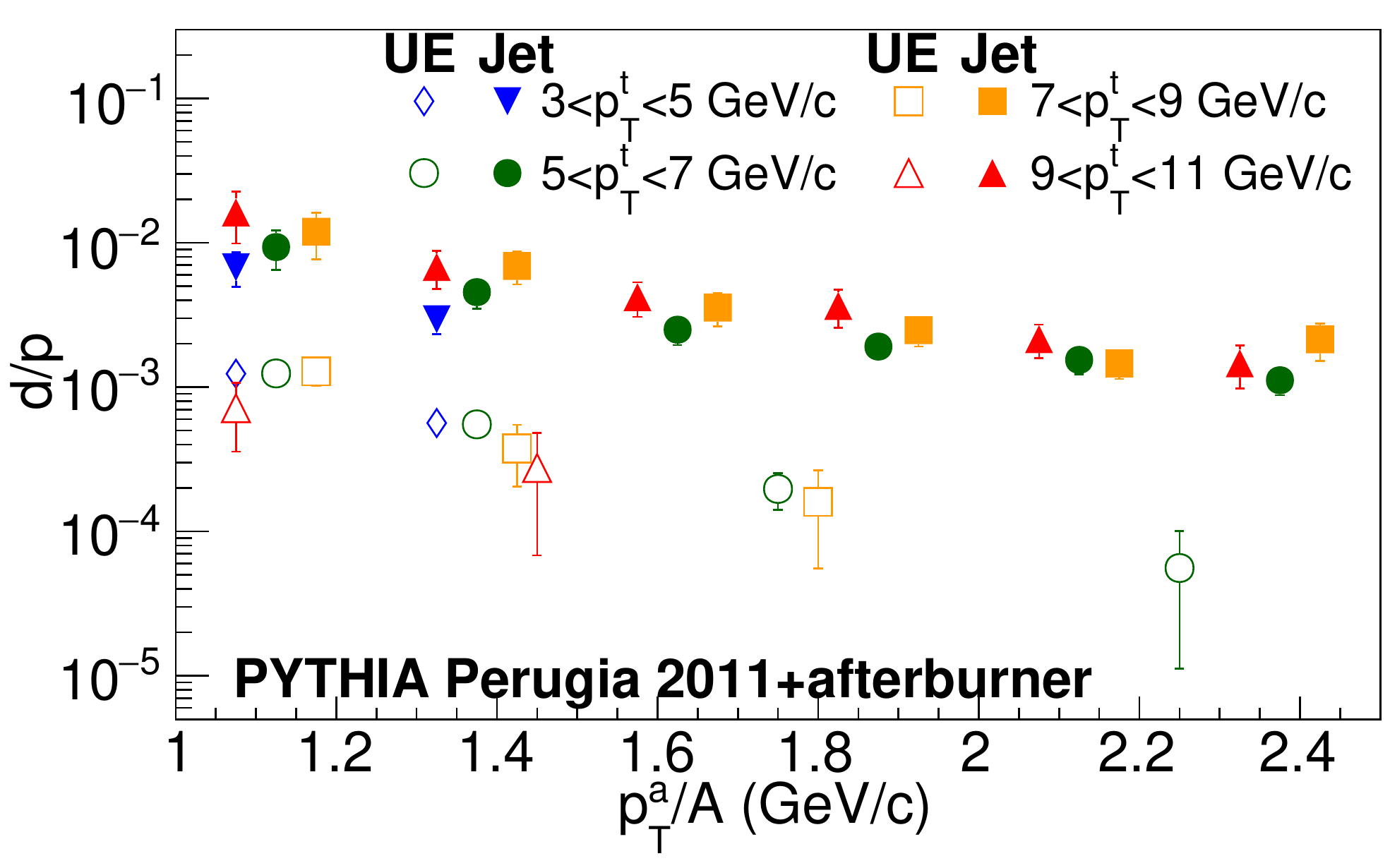}
}}\caption{Ratio of  deuteron to proton yields  as a function of $p_T^a/A$ calculated from the conditional spectra for \pttrigrange{3}{5}, \pttrigrange{5}{7}, \pttrigrange{7}{9}, and \pttrigrange{9}{11} for both the jet-like correlation and the underlying event.  Points have been displaced for visibility.}\label{Fig:ConditionalRatioPYTHIA}
\end{center}
\end{figure}

Figure~\ref{Fig:ConditionalRatioPYTHIA} shows the ratio of deuteron to proton yields
plotted as a function of $p_T^a/A$ where $A$ is the mass number. The ratio is 
calculated from the conditional spectra
using following equation
\begin{equation}
\frac{d}{p} = \frac{  \frac{1}{2\pi (p_{T}^{d}/2)}\frac{dY_{trig}^d}{dp_{T}^{d}/2}  }{  \frac{1}{2\pi p_{T}^{p}}\frac{dY_{trig}^p}{dp_{T}^{p}} }
\end{equation}
 for \pttrigrange{3}{5}, \pttrigrange{5}{7}, \pttrigrange{7}{9}, and \pttrigrange{9}{11} for both the jet-like correlation and the underlying event.  The ratio for  jet-like correlation is higher than that for the underlying event. The ratio remains similar for different trigger \pT for both the underlying event and jet-like correlations and decreases as a function of  $p_T^a/A$. However, the ratio decreases faster for the underlying event with increasing $p_T^a/A$, suggesting the contribution to the $d/p$ ratio from jet-like contributions is greater at higher \pT.

\begin{figure}
\begin{center}
\rotatebox{0}{\resizebox{\columnwidth}{!}{
        \includegraphics{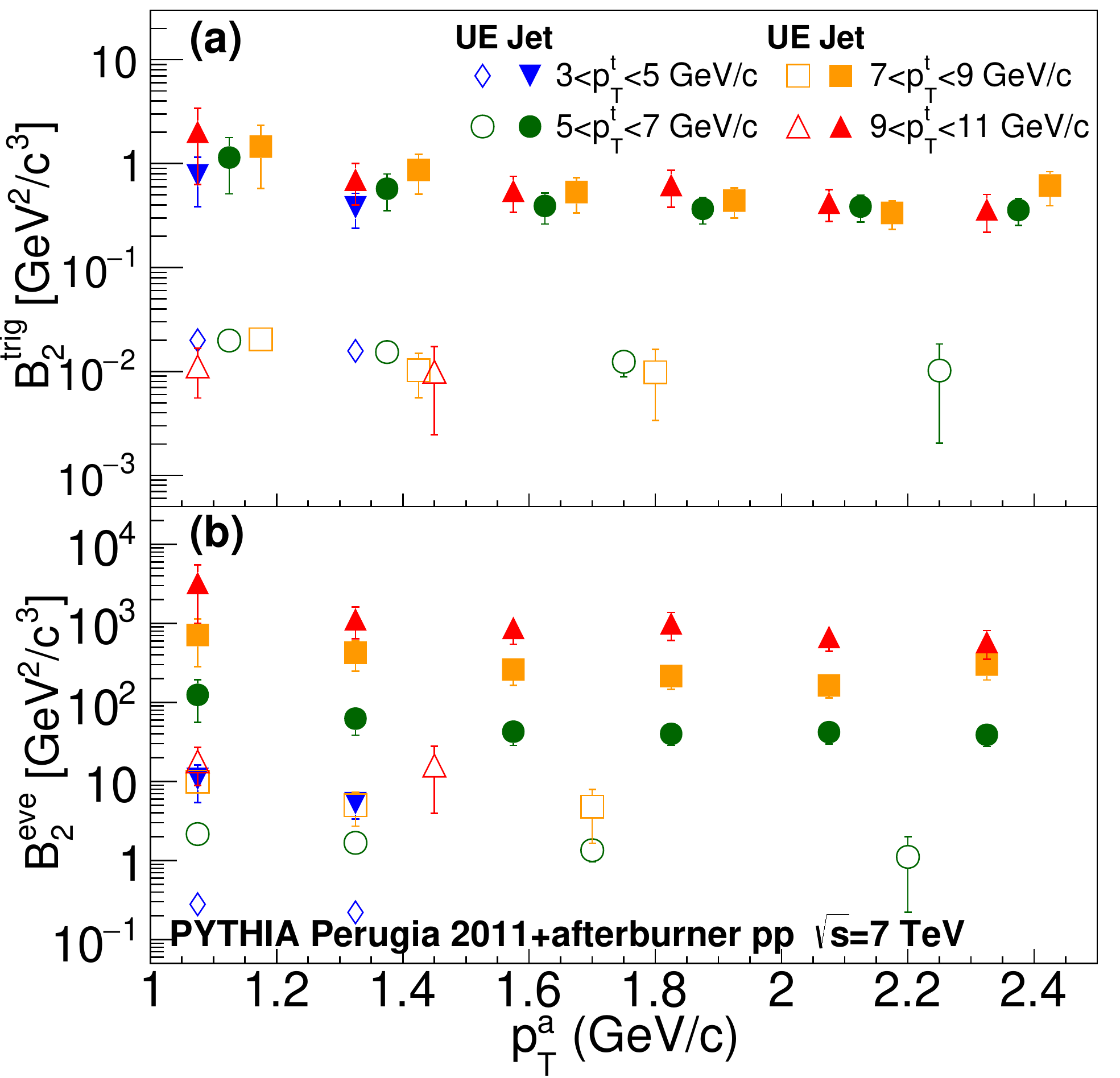}
}}\caption{The (a) \Btwotrig and (b) \Btwoeve calculated from the conditional yields for \pttrigrange{3}{5}, \pttrigrange{5}{7}, \pttrigrange{7}{9}, and \pttrigrange{9}{11} for both the jet-like correlation and the underlying event.  Points have been displaced for visibility.}\label{Fig:ConditionalB2PYTHIA}
\end{center}
\end{figure}

An analog to \Btwo can be calculated for these conditional spectra
\begin{equation}
 B_2^{trig} = \frac{  \frac{1}{2\pi (p_{T}^{d}/2)}\frac{dY_{trig}^d}{dp_{T}^{d}/2}  }{  (\frac{1}{2\pi p_{T}^{p}}\frac{dY_{trig}^p}{dp_{T}^{p}})^2 },
\end{equation}
where the superscript denotes the normalization by the number of triggers.  An analogous quantity for normalization by the number of events can be calculated, \Btwoeve, where $Y_{trig}$ is replaced by $Y_{eve}$.  

\Fref{Fig:ConditionalB2PYTHIA} shows the \Btwotrig and \Btwoeve from the conditional yields with both normalizations for several trigger momenta.  
The \Btwotrig for both the jet-like correlation and the underlying event are comparable for all trigger momenta.  Since \Btwotrig is a measure of the particle composition of jets, this indicates that the deuteron to proton ratio is independent of the trigger momenta.  For both the jet-like correlation and the underlying event, \Btwotrig and \Btwoeve are roughly independent of $p_T^a/T$.

\Fref{Fig:ConditionalB2PYTHIA}(b) shows \Btwoeve, which increases with increasing \pttrig.  This shows that the relative contribution of a jet to inclusive \Btwo increases with the jet momentum.
The background for these correlations contain contributions not only from soft processes but also from hard processes unrelated to the trigger particle.  At low momenta, several hard processes can occur per event. 
The increase in \Btwoeve may therefore occur from several hard processes in the event.  It may also occur due to a higher overall multiplicity in events with a high momentum hadron.

Comparing 
Fig.~\ref{Fig:ConditionalB2PYTHIA} 
to 
Fig.~\ref{Fig:B2PYTHIA} 
suggests that \Btwo as a function of \pT in Fig.~\ref{Fig:B2PYTHIA} may be dominated by contributions from the underlying event at lower \pT and for higher \pT the contribution from jet-like correlations dominates. This may suggest that deuteron production at high \pT is mostly due to jets.

\section{Conclusions}\label{Sec:Conclusions}
 We modeled the production of deuterons through coalescence in \pp collisions at \sqrts = 7 TeV by applying an afterburner which coalesces protons and neutrons in PYTHIA. 
This model overpredicts the data, which is likely because PYTHIA uses the primary vertex as the origin of primary particles and therefore likely overestimates spatial correlations between protons and neutrons.  However, the shape of proton and deuteron spectra and hence \Btwo as a function of \pT in the model is roughly consistent with the data.
These calculations show that the rise in \Btwo with momentum can indeed be generated by deuteron formation through coalescence in jets. 
We then showed that \dhcs can be used to separate deuterons in jets from those in the underlying event. This leads to predictions that the conditional \Btwotrig should be roughly independent of the trigger particle momentum if deuterons are produced through coalescence in jets.  This approach may also be useful in heavy ion collisions, where correlations between protons and neutrons can arise both through the production of jets and through hydrodynamical flow.   \section{Acknowledgements}
 We are grateful to Paul Stankus for useful comments in the manuscript.  This work was supported in part by funding from the Division of Nuclear Physics of the U.S. Department of Energy under Grant No. DE-FG02-96ER40982. N.S. acknowledges the support of DST-SERB Ramanujan Fellowship (D.O. No. SB/S2/RJN- 269084/2015). LK acknowledges the support of the SERB grant No. ECR/2016/000109.  This material is based upon work performed using computational resources within the University of Tennessee (UT) Advanced Computing Facility (http://www.nics.utk.edu/computing-resources/acf) that are supported by the UT Joint Institute for Computational Sciences (http://www.jics.utk.edu) and partners. Any opinions, findings, and conclusions or recommendations expressed in this material are those of the author(s) and do not necessarily reflect the views of the University of Tennessee, the Joint Institute for Computational Sciences, or other ACF partners. 
\bibliography{Bibliography,b2_paper}   

\end{document}